\documentclass[twocolumn]{aastex6}

\slugcomment{Accepted for publication in the Astrophysical Journal}

\shorttitle{Dust in a normal galaxy at $z>4$}
\shortauthors{Pope et al.}

\fullcollaborationName{The Friends of AASTeX Collaboration}

\begin{document}


\title{Early Science with the Large Millimeter Telescope: Detection of dust emission in multiple images of a normal galaxy at $z>4$ lensed by a Frontier Fields cluster}


\author{Alexandra Pope\altaffilmark{1}, Alfredo Monta\~{n}a\altaffilmark{2,3}, Andrew Battisti\altaffilmark{1}, Marceau Limousin\altaffilmark{4}, Danilo Marchesini\altaffilmark{5}, Grant W.~Wilson\altaffilmark{1}, Stacey Alberts\altaffilmark{6}, Itziar Aretxaga\altaffilmark{2}, Vladimir Avila-Reese\altaffilmark{7}, Jos\'e Ram\'on Bermejo-Climent\altaffilmark{15}, Gabriel Brammer\altaffilmark{8}, Hector Bravo-Alfaro\altaffilmark{9}, Daniela Calzetti\altaffilmark{1}, Ranga-Ram Chary\altaffilmark{10}, Ryan Cybulski\altaffilmark{1,5}, Mauro Giavalisco\altaffilmark{1}, David Hughes\altaffilmark{2}, Erin Kado-Fong\altaffilmark{5}, Erica Keller\altaffilmark{11}, Allison Kirkpatrick\altaffilmark{1,14}, Ivo Labbe\altaffilmark{12}, Daniel Lange-Vagle\altaffilmark{5}, James Lowenthal\altaffilmark{13}, Eric Murphy\altaffilmark{11}, Pascal Oesch\altaffilmark{14}, Daniel Rosa Gonzalez\altaffilmark{2}, David S\'anchez-Arg\"uelles\altaffilmark{2}, Heath Shipley\altaffilmark{5}, Mauro Stefanon\altaffilmark{12}, Olga Vega\altaffilmark{2}, Katherine Whitaker\altaffilmark{1,16}, Christina C.~Williams\altaffilmark{6}, Min Yun\altaffilmark{1}, Jorge A.~Zavala\altaffilmark{2}, Milagros Zeballos\altaffilmark{2,17}
}


\altaffiltext{1}{Department of Astronomy, University of Massachusetts, Amherst, MA 01003; pope@astro.umass.edu}
\altaffiltext{2}{Instituto Nacional de Astrof\'{i}sica, \'{O}ptica y Electr\'{o}nica (INAOE), Luis Enrique Erro 1, Sta. Ma. Tonantzintla, 72840 Puebla, Mexico} 
\altaffiltext{3}{Consejo Nacional de Ciencia y Tecnolog\'{i}a (CONACYT), Av. Insurgentes Sur 1582, Col. Cr\'{e}dito Constructor, Del. Benito Ju\'{a}rez, C.P.: 03940, D.F., M\'exico} 
\altaffiltext{4}{Aix Marseille Univ, CNRS, LAM, Laboratoire dÕAstrophysique de Marseille, Marseille, France}
\altaffiltext{5}{Department of Physics and Astronomy, Tufts University, Medford, MA 02155, USA}
\altaffiltext{6}{Steward Observatory, University of Arizona, 933 North Cherry Avenue, Tucson, AZ 85721, USA}
\altaffiltext{7}{Instituto de Astronom\'ia, Universidad Nacional Aut\'onoma de M\'exico,  A.P. 70-264, 04510, CDMX}
\altaffiltext{8}{Space Telescope Science Institute, 3700 San Martin Dr., Baltimore, MD 21218, USA} 
\altaffiltext{9}{Departamento de Astronomia, Universidad de Guanajuato, Apdo. Postal 144, Guanajuato 36000. Mexico} 
\altaffiltext{10}{Infrared Processing and Analysis Center, MS314-6, California Institute of Technology, Pasadena, CA 91125, USA}
\altaffiltext{11}{National Radio Astronomy Observatory, 520 Edgemont Road, Charlottesville, VA 22903, USA} 
\altaffiltext{12}{Leiden Observatory, Leiden University, NL-2300 RA Leiden, The Netherlands} 
\altaffiltext{13}{Dept. of Astronomy, Smith College, Northampton, MA  01063}
\altaffiltext{14}{Yale Center for Astronomy and Astrophysics, Astronomy Department, Yale University, New Haven, CT 06511, USA}
\altaffiltext{15}{Departamento de Astrof\'{i}sica, Universidad de La Laguna. V'a L\'{a}ctea s/n, La Laguna 38200, Tenerife, Spain}
\altaffiltext{16}{Department of Physics, University of Connecticut, Storrs, CT 06269, USA}
\altaffiltext{17}{Instituto Tecnol\'{o}gico Superior de Tlaxco, Predio Cristo Rey Ex-Hda de Xalostoc s/n, 90250 Tlaxcala, Mexico}

\begin{abstract}

We directly detect dust emission in an optically-detected, multiply-imaged galaxy lensed by the Frontier Fields cluster MACSJ0717.5+3745. We detect two images of the same galaxy at 1.1$\,$mm with the AzTEC camera on the Large Millimeter Telescope leaving no ambiguity in the counterpart identification. This galaxy, MACS0717\_Az9, is at $z>4$ and the strong lensing model ($\mu=7.5$) allows us to calculate an intrinsic IR luminosity of $9.7\times10^{10}L_{\odot}$ and an obscured star formation rate of $14.6\pm4.5\,\rm{M_{\odot}/yr}$. The unobscured star formation rate from the UV is only $4.1\pm0.3\,\rm{M_{\odot}/yr}$ which means the total star formation rate ($18.7\pm4.5\,\rm{M_{\odot}/yr}$) is dominated (75--80\%) by the obscured component. 
With an intrinsic stellar mass of only $6.9\times10^{9}\,\rm{M_{\odot}}$, MACS0717\_Az9 is one of only a handful of $z>4$ galaxies at these lower masses that is detected in dust emission. This galaxy lies close to the estimated star formation sequence at this epoch. However, it does not lie on the dust obscuration relation (IRX-$\beta$) for local starburst galaxies and is instead consistent with the Small Magellanic Cloud (SMC) attenuation law. 
This remarkable lower mass galaxy showing signs of both low metallicity and high dust content may challenge our picture of dust production in the early Universe. 

\end{abstract}




\section{Introduction} \label{sec:intro}

Over the past 20 years, surveys at rest-frame UV wavelengths have mapped out the history of unobscured star formation from the present day back to $z\sim8$ (Madau \& Dickinson 2014). However, the roughly equal brightness of the cosmic infrared and optical backgrounds informs us that half the light from the formation and evolution of galaxies is obscured by dust (Lagache et al.~2005). Surveys with the {\it Spitzer Space Telescope} and the {\it Herschel Space Observatory} showed that the contribution from infrared-luminous galaxies to the star formation rate density increases dramatically from $z=0$--2 (e.g.~Caputi et al.~2007; Murphy et al.~2011a; Magnelli et al.~2013). Beyond $z\sim3$, our census of the dust-obscured, and hence {\em total}, star formation activity is severely incomplete.  

Until recently, surveys of dust-obscured activity at $z>3$ detected only the bright ultra-luminous infrared galaxies (ULIRGs, $L_{\rm{IR}}>10^{12}L_{\odot}$, Casey et al.~2014). While ULIRGs are prevalent at high redshift and many are not extreme starbursts like their local counterparts, they are not responsible for creating the bulk of the stars in the Universe (Lagache et al.~2005). At $z\sim2$, much of the cosmic star formation activity is occurring in galaxies with $L_{\rm{IR}}<10^{12}L_{\odot}$ (Murphy et al.~2011a; Magnelli et al.~2013). While these normal\footnote{We use ``normal" to refer to typical star-forming galaxies for their epoch; on the star formation sequence (Noeske et al.~2007), and/or with stellar masses near the knee of the stellar mass function (Muzzin et al.~2013).} galaxies can be selected at UV wavelengths, we have yet to directly detect most of their star formation activity as it is obscured by dust. The UV slope can provide an estimate of the dust extinction in the local Universe (Meurer et al.~1999); however, this correction is uncertain at high redshift where star formation is clumpy (Guo et al.~2015)
and gas and dust are more widely distributed across the galaxy (e.g.~Ivison et al.~2011).

With its exceptional sensitivity, ALMA can directly detect dust in normal galaxies out to and beyond $z\sim3$. In ALMA Cycles 0-2, several programs have pushed below the ULIRG limit, detecting dust in half a dozen UV-selected galaxies from $z=4$--7.5 (e.g.~Capak et al.~2015; Watson et al.~2015; Willott et al.~2015; Dunlop et al.~2017). These studies show mixed results with some sources having significant dust emission while others remain undetected (e.g.~Schaerer et al.~2015; Bouwens et al.~2016). 

A complementary facility for directly detecting dust in $z>4$ galaxies is the Large Millimeter Telescope Alfonso Serrano (LMT, Hughes et al.~2010). With a large aperture and fast mapping capability, the AzTEC camera (Wilson et al.~2008) on the 32m LMT can survey dust in galaxies down to $L_{\rm{IR}}\sim6\times10^{11}L_{\odot}$ regardless of redshift due to the negative K-correction. Gravitational lensing can be used to push even deeper. In this paper, we present the direct detection of dust in a multiply-imaged normal galaxy at $z>4$ with AzTEC on LMT. 

Throughout this paper we assume a cosmology with $H_{0}=70\,\rm{km}\,\rm{s}^{-1}\,\rm{Mpc}^{-1}$, $\Omega_{\rm{M}}=0.3$, and $\Omega_{\Lambda}=0.7$.

\section{Data}
\label{sec:data}

\subsection{Frontier Fields Program}

The Frontier Fields (FFs) program\footnote{http://www.stsci.edu/hst/campaigns/frontier-fields/} started as a large {\it Hubble Space Telescope} ({\it HST}) survey of low redshift clusters in order to identify and study high redshift background galaxies that are gravitationally lensed. In this paper, we use the 13-band {\it HST} data, the {\it Spitzer}-IRAC imaging from 3.6 to 8 micron, and $K$-band imaging from Keck-MOSFIRE (program N097M and N135M, PI: Marchesini, Brammer et al.~2016). The {\it HST} data include the F435W, F606W, F814W, F105W, F125W, F140W, and F160W images from the FF program; the F475W, F625W, F775W, and F850LP images from CLASH (Postman et al. 2012); and the F275W and F336W images from the program GO-13389 (PI: Siana). 
The v2.1 UV-to-IRAC multi-wavelength photometric catalog used in this paper was constructed following Skelton et al. (2014). The final catalog construction accounting for the intra-cluster light and contamination from brightest cluster galaxies will be described in Shipley et al. (2017).

Since our target is a multiply-imaged, strongly-lensed galaxy, interpretation of its intrinsic properties will depend on the lensing model. 
STScI has released magnification maps as a function of background galaxy redshift for all FF clusters calculated from several independent lensing models\footnote{https://archive.stsci.edu/prepds/frontier/lensmodels/}. In this paper, we use the updated lensing models from Limousin et al.~(2016) and Diego et al.~(2015), and we verified that our results are robust with other lensing models from STScI (Johnson et al.~2014; Zitrin et al.~2015). We present our results for two different lensing models to give a sense of how the parameters we are interested in (stellar mass, star formation rate, UV slope) change under different lensing models. 

\begin{figure*}[ht!]
\plottwo{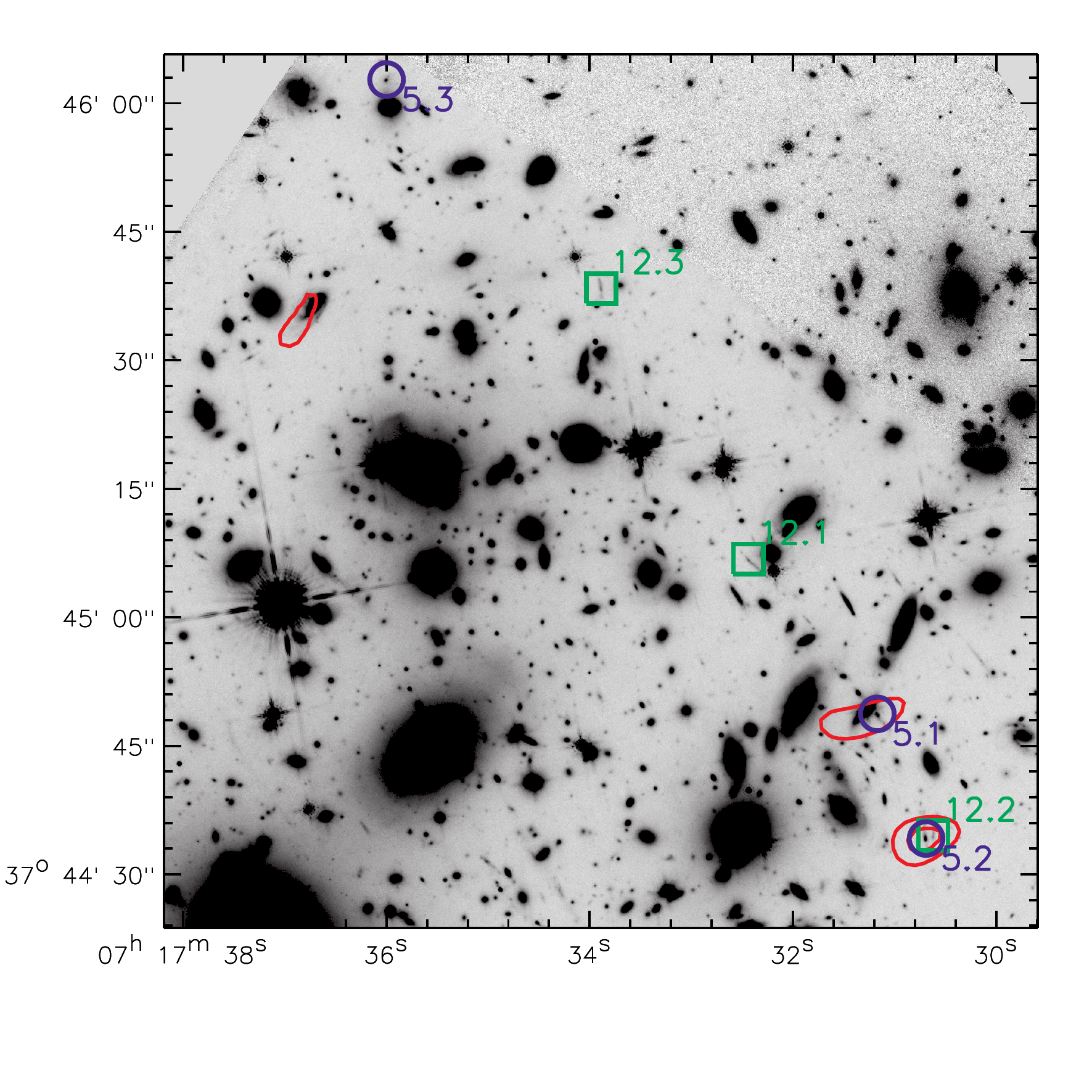}{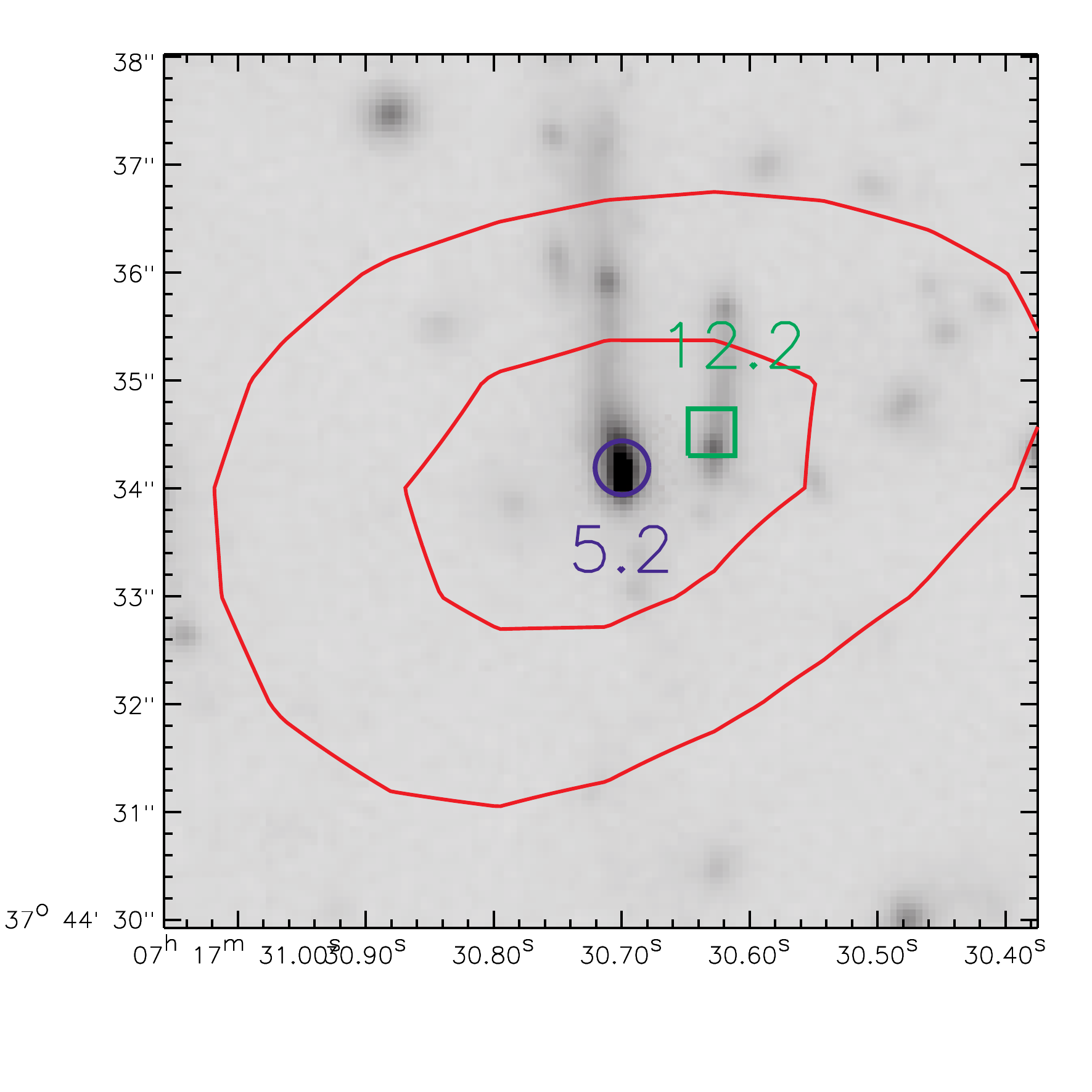}
\caption{{\it HST} F160W image towards MACSJ0717 with AzTEC/LMT contours (3 and $3.5\sigma$) in red. Our AzTEC map covers a wider region than shown here. The $3\sigma$ detection in the top left is unrelated to the multiply-imaged galaxy. We show the two multiply-imaged systems: 5.1/5.2/5.3 at $z>4$ (blue circles) and 12.1/12.2/12.3 at $z=1.71$ (green squares). The {\it right} panel shows a zoom-in of the millimeter detection MACS0717\_Az9, which is at the bottom right of the {\it left} panel. The size of the image is roughly equal to the FWHM of the beam.
}
\label{fig:hst}
\end{figure*}

\subsection{New AzTEC/LMT observations}
\label{sec:az}

In November and December 2014, we imaged the FF cluster MACSJ0717.5+3745 with AzTEC during early science with the LMT. During early science operations, we are using the inner 32m of the eventual 50m aperture\footnote{The LMT is transitioning to a 50m telescope in 2017.}.  
AzTEC is a 1.1$\,$mm bolometer array camera, with a beam size of 8.5 arcsec (FWHM) on the 32$\,$m LMT. 
Data were taken in good weather conditions ($\tau_{\rm{225GHz}}=0.05$--0.12). The on-source integration time was 21.1 hours. Our map covers 25 sq.~arcmin field reaching a mean RMS of 0.24$\,$mJy (central RMS is 0.19$\,$mJy). 

The calibration and analysis of the AzTEC data follow the procedure described in Wilson et al.~(2008) and Scott et al.~(2008).
The results on the number counts and source properties from the full LMT FF program\footnote{http://people.umass.edu/apope1/FF/} 
will be presented in future papers. Here we focus on a unique and rare source detected in our AzTEC map (MACS0717\_Az9), which is coincident with an optically-detected, multiply-imaged lensed galaxy (known as 5.1/5.2/5.3, Zitrin et al.~2009). 
This is the only strongly-lensed, multiply-imaged system detected in our AzTEC survey. Figure \ref{fig:hst} {\it left} shows AzTEC contours on the {\it HST} F160W image; two optical images (labelled 5.2 and 5.1) of the known multiply-imaged system are detected with AzTEC (3.7 and 3.3$\sigma$ respectively). In Section \ref{sec:cpt}, we demonstrate that at least half the millimeter flux detected by AzTEC must be associate with this system. 
A third $>3\sigma$ AzTEC detection is visible in the top-left corner of Figure \ref{fig:hst} {\it left}, but it is unassociated with the multiply-imaged system that is the focus of this paper.

\subsection{Robustness of millimeter detections}
\label{sec:robust}

In order to test the robustness of MACS0717\_Az9 both as a millimeter source and as the counterpart to the $z>4$ multiply-imaged galaxy, we perform several simulations. We stress that since we have prior information on the positions of a known multiply-imaged galaxy, we have more confidence in lower signal-to-noise detections.

First we determine the chance that our millimeter detections (3.7 and 3.3$\sigma$) of the multiple-images 5.2 and 5.1 are spurious.
We perform source extraction on 3000 noise maps. As with our original source list, we limit source extraction to regions of the map with noise $<0.4\,$mJy. We detect an average of 1 and 4 random $>3.7\sigma$ and $>3.3\sigma$ sources respectively in the noise maps. However, the sources we are interested in are not at random positions and specifically we detect two components of a previously known multiply-imaged system. With this prior information, we find the chance of randomly detecting a $>3.7\sigma$ source within 1 AzTEC beam of 5.2 and a $>3\sigma$ source within 1 AzTEC beam of 5.1 is $<0.03\%$. We find the same answer if we vary the position of 5.2 and 5.1 on the map but conserve their relative separation. Therefore, the probability that our millimeter detection of this multiply-imaged system (5.1 and 5.2) is a spurious detection is negligible.

Second, we test the chance that we should detect a multiply-imaged source given the known lensing models and a model for a background population of millimeter sources.
We develop 500 simulated maps using the empirical galaxy evolution model of Bethermin et al. (2012) for the input background millimeter galaxies and the lensing models of MACSJ0717 from the CATS group (Limousin et al. 2016). Here we report results using the {\it cored} mass model, and verified that there are no measurable difference with the {\it non-cored} mass model. 
We randomly populate simulated maps with millimeter sources down to intrinsic (e.g.~non-lensed) 1.1mm fluxes of $\ge0.01\,$mJy. Redshifts are assigned to each millimeter source from the Bethermin et al.~(2012) model.
Then we run the background population through the cluster using the mass model in LENSTOOL (Kneib et al.~1996; Jullo et al.~2007) to find the observed population of millimeter sources and ask how often the millimeter sources are multiply-imaged. With no observed flux limit, multiple-image systems are found in all (99.4\%) of our simulated maps, with an average of 6-7 systems per map. When we impose an observed flux limit of $0.7\,$mJy (i.e. $>3.5\sigma$), we find that 30\% of these multiple systems have at least one image detectable in our simulated maps. 
Coupled with our estimated completeness limit of 50\% at this low flux level (Monta\~{n}a et al.~in prep.), our simulations predict that we will detect 1 multiply-imaged system in our AzTEC map of MACS0717. 
Besides MACS0717\_Az9, there are no other known multiply-imaged sources in MACS0717 (using catalogs of known multiply-imaged systems, e.g~Limousin et al.~2016) that are individually detected in our AzTEC maps. We take the full list of multiply-imaged sources including their magnification values and we stack the intrinsic millimeter flux for each multiply-imaged source. We do not find that any other systems are detected even when averaging the individual components in this way.
Furthermore, none of the other AzTEC detections are in regions of strong magnification. Therefore, our simulations predict that we should detect one multiply-imaged systems like MACS0717\_Az9. 

Finally, we can further test the robustness of this millimeter detection by showing that the millimeter fluxes that we measure for 5.1, 5.2 and 5.3 are consistent with each other given their known magnifications (see Section \ref{sec:cpt}). The results of all three of these simulations and tests show that we have unambiguously detected dust emission in this multiply-imaged system.

\section{Analysis}

\begin{figure*}[ht!]
\vspace{-0.9in}
\plottwo{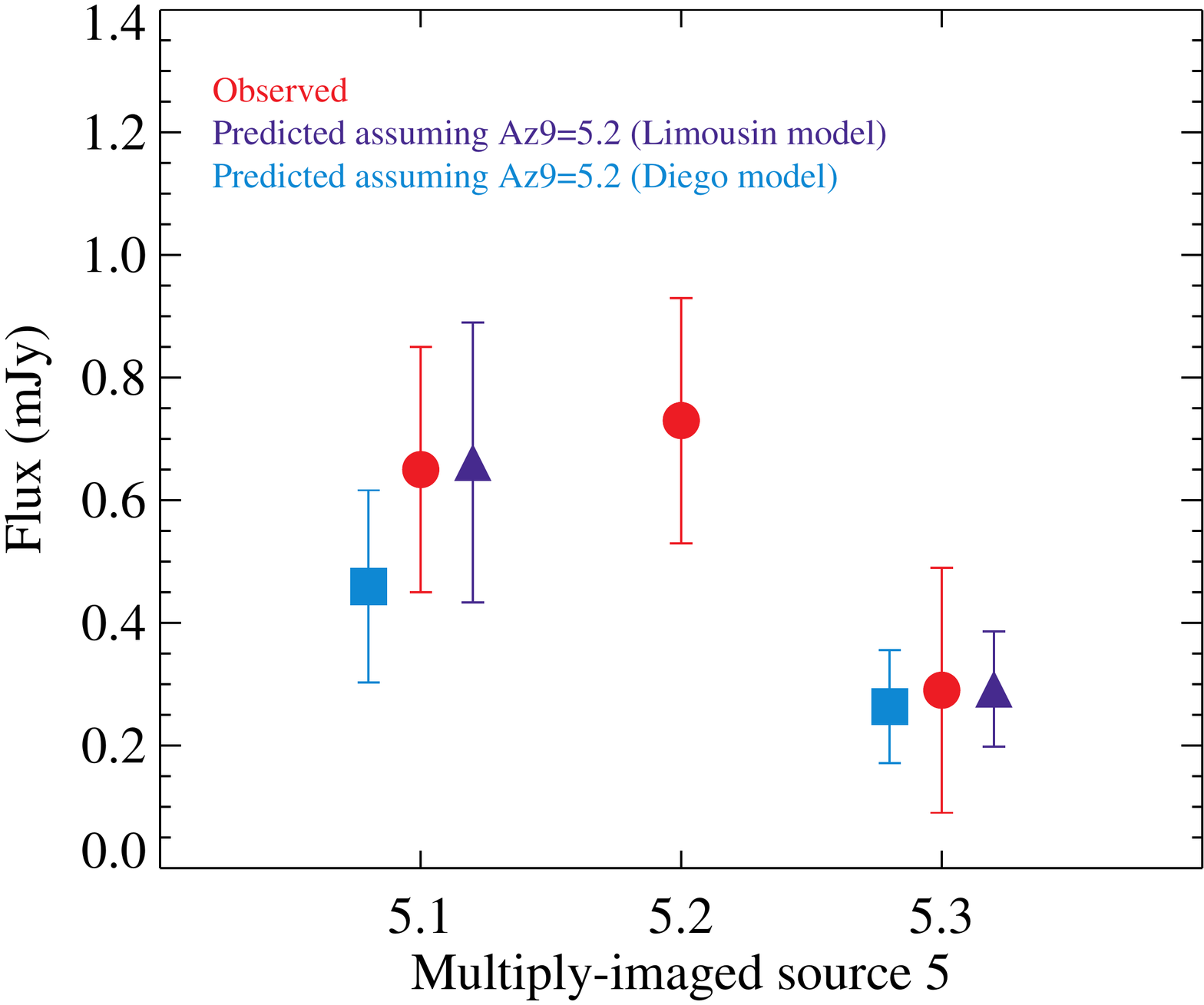}{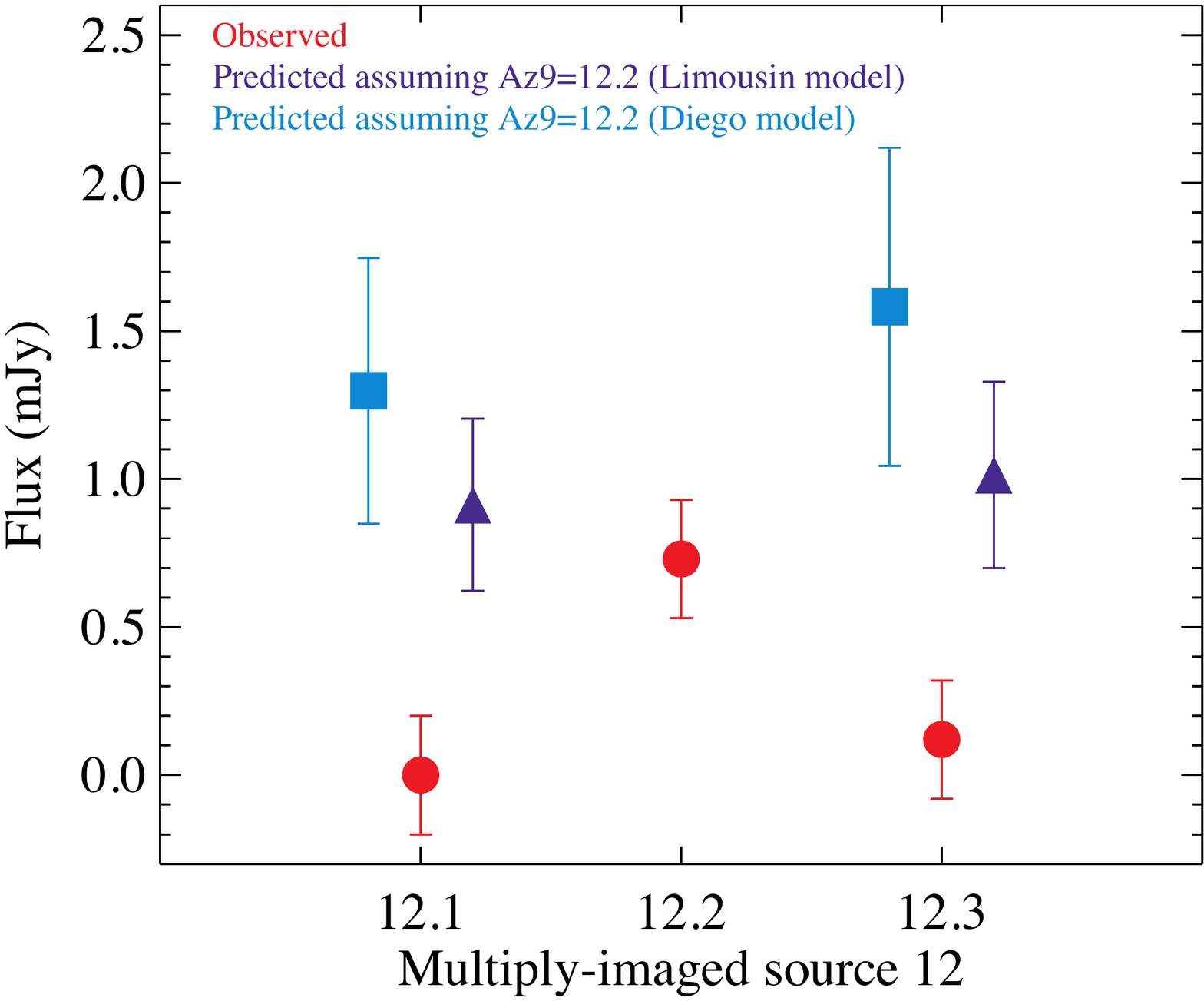}
\vspace{-0.35in}
\caption{Demonstration that MACS0717\_Az9 is most likely associated with the multiply-imaged galaxy 5.2. ({\it left}) Red circles show the observed 1.1$\,$mm fluxes of the 3 multiply-imaged components; 5.1, 5.2 and 5.3, from our AzTEC maps. The dark blue triangles and light blue squares show the predicted millimeter fluxes of 5.1 and 5.3 assuming that MACS0717\_Az9 is associated with 5.2 and applying the known lensing magnifications (Table \ref{tab:red}) from the Limousin et al.~(2016) and Diego et al.~(2015) lensing models, respectively. The error bars on the blue points include the photometric uncertainty and the magnification uncertainty. 
({\it right}) Same as the {\it left} panel but for the multiply-imaged source 12.1/12.2/12.3 showing the observed fluxes are inconsistent with the predicted fluxes under these two different lensing models. 
}
\label{fig:flux}
\end{figure*}

\subsection{Counterpart of MACS0717\_Az9}
\label{sec:cpt}
Before we can discuss the nature of this millimeter source, we need to demonstrate that the multiply-imaged system is the correct optical counterpart. From our simulations (Monta\~{n}a et al.~in prep.), we find a positional accuracy for this system of 3.1 arcsec with 90\% confidence.  
Within the search radius of MACS0717\_Az9, we find two multiply-imaged systems (Figure \ref{fig:hst} {\it right}): 5.2 ($z>4$) and 12.2 ($z=1.71$). 
But our AzTEC map also covers the other multiple images of these systems: 5.1/5.3 and 12.1/12.3 (Figure \ref{fig:hst} {\it left}). 
We detect 5.1 in our AzTEC map at $3.3\sigma$. Source 5.3 has a lower magnification; as a result, the measured AzTEC flux is only $1.4\sigma$ (Table \ref{tab:obs}). Both 12.1 and 12.3 are undetected in our AzTEC map. In this Section, we show that the counterpart to MACS0717\_Az9 must be 5.2. 

The magnifications ($\mu$) are known for all components of systems 5 and 12 (Limousin et al.~2016; Diego et al.~2015), and so we test if the observed fluxes of each component are consistent with the predicted fluxes. Figure \ref{fig:flux} {\it left} shows the observed fluxes of all three images of system 5 as the solid circles. If we assign some fraction $F$ of the MACS0717\_Az9 flux to 5.2 ($S_{\rm{5.2,obs}}=S_{\rm{Az9}}*F$) and the remainder of the flux to 12.2, then we can predict the flux of 5.1 and 5.3 as follows: 

\begin{eqnarray}
S_{\rm{5.1,pred}}=(S_{\rm{5.2,obs}}/\mu_{5.2})\times \mu_{5.1}\\
S_{\rm{5.3,pred}}=(S_{\rm{5.2,obs}}/\mu_{5.2})\times \mu_{5.3}
\end{eqnarray}

\noindent where $\mu_{5.2}$, $\mu_{5.1}$ and $\mu_{5.3}$ denote the known magnifications of 5.2, 5.1 and 5.3. 
The triangles and squares in Figure \ref{fig:flux} {\it left} show the predicted fluxes from two different lensing models assuming $F=1$, which are remarkably consistent with the observed fluxes of 5.1 and 5.3. 
If instead we perform this calculation assuming the millimeter emission comes from 12.2 (Figure \ref{fig:flux} {\it right}), we find that the observed fluxes of 12.1 and 12.3 are inconsistent with the expected fluxes under each of the two lensing models. 

\begin{figure}[ht!]
\includegraphics[trim=0 0 0 0,clip,scale=0.85]{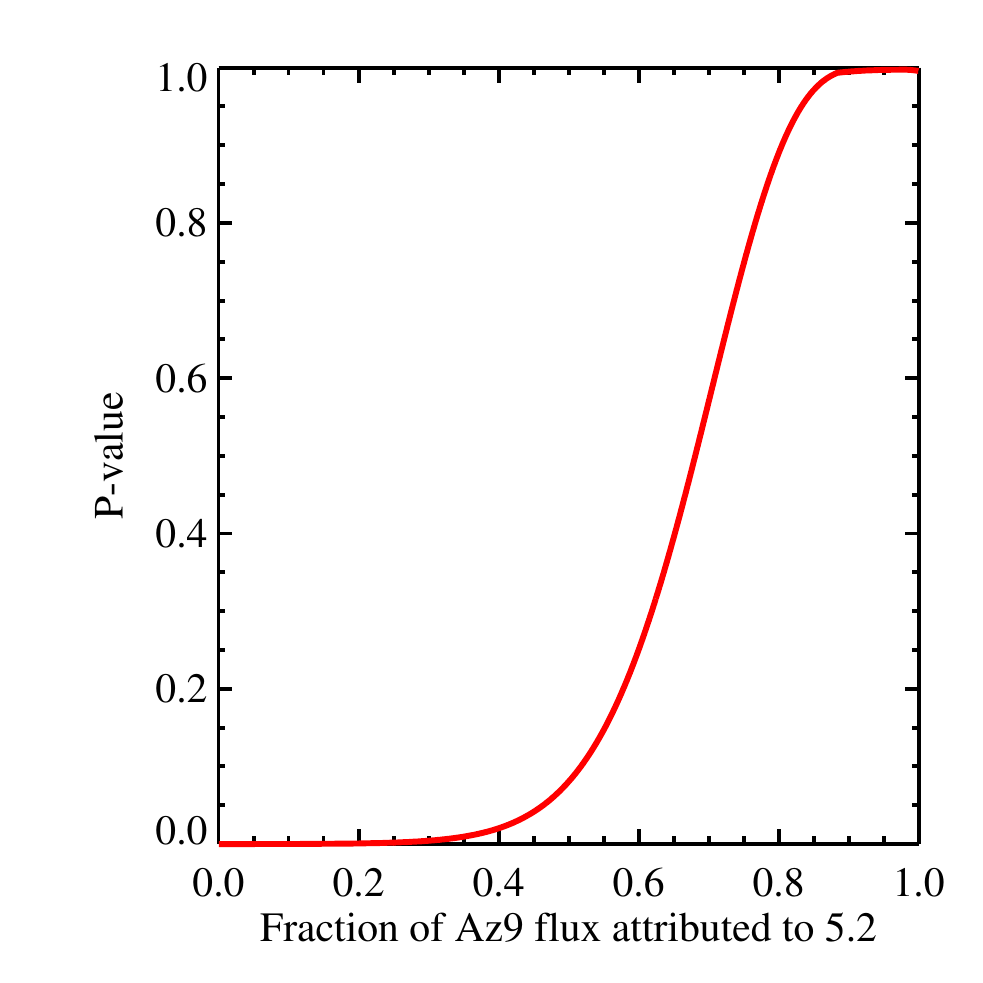}  
\caption{P-value from testing the hypothesis that the observed fluxes of 5.1/5.3 and 12.1/12.3 are consistent with the predicted fluxes as a function of the fraction of the flux from MACS0717\_Az9 that is attributed to 5.2. 
The most likely scenario is that all of the flux from MACS0717\_Az9 is associated with 5.2, and we can rule out the scenario where $\le45\%$ of the millimeter flux is coming from 5.2.
}
\label{fig:Pval}
\end{figure}

Next, we calculate the $P$-value for all values of $F$ from 0--1 under the hypothesis that the predicted fluxes equal the observed fluxes:  

\begin{eqnarray}
S_{\rm{5.1,obs}}-S_{\rm{5.1,pred}} \sim \rm{N}(0,\sigma_{5.1,obs}^{2}+\sigma_{5.1,pred}^{2})\\
S_{\rm{5.3,obs}}-S_{\rm{5.3,pred}} \sim \rm{N}(0,\sigma_{5.3,obs}^{2}+\sigma_{5.3,pred}^{2})
\end{eqnarray}

\noindent where N is a normal distribution. $\sigma_{5.1,pred}$ and $\sigma_{5.3,pred}$ include the uncertainties from all quantities in Equations 1 and 2; the flux measurement of 5.2, the magnification of 5.2 and the magnification of 5.1 and 5.3, respectively. We perform this hypothesis test for 5.1, 5.3, 12.1 and 12.3 and combine the $P$-values using Fisher's method (Fisher 1925). The combined $P$-value as a function of $F$ is plotted in Figure \ref{fig:Pval}. We can reject the null hypothesis that $F\le0.45$ at a significance level of 0.05; this means that at least half the flux of MACS0717\_Az9 {\it must} be associated with 5.2. For the analysis in this paper we assume the most likely scenario: that all of the flux of MACS0717\_Az9 is associated with 5.2  (i.e.,~$F=1$). In Section \ref{sec:whatif}, we discuss how our main results are affected under the conservative assumption that only half the millimeter flux is associated with 5.2.

\begin{table*}
\begin{center}
\scriptsize
\caption{Observed properties of dust emission in the multiply-imaged system 5.1/5.2/5.3 in MACSJ0717.5+3745}
\label{tab:obs}
\begin{tabular}{lllllll}
\hline
ID & z$^{\rm{a}}$   & Observed $S_{1.1\rm{mm}}$$^{\rm{b}}$ & Intrinsic $S_{1.1\rm{mm}}$$^{\rm{c}}$ & $L_{\rm{IR}}$$^{\rm{d}}$ \\
 &      & (mJy)  & (mJy) & ($10^{10}\,L_{\odot}$) \\\hline
5.1          & 	4.1 &  $0.65\pm0.20$ &  $0.096\pm0.033$ & $9.6\pm4.2$  \\
5.2 (MACS0717\_Az9)       & 	4.1  & $0.73\pm0.20$ &  $0.097\pm0.029$ & $9.7\pm3.9$ \\ 
5.3        &   	4.1& $0.29\pm0.21$ &  $0.097\pm0.067$ & $9.7\pm7.2$  \\\hline
5 (average)        	 & 	&&  $0.097\pm0.026$ & $9.7\pm3.0$  \\\hline\hline
5.1         &  	4.3 & $0.65\pm0.20$ &   $0.144\pm0.048$ & $14.4\pm6.2$  \\
5.2 (MACS0717\_Az9)        & 	4.3 &$0.73\pm0.20$ &  $0.102\pm0.032$ & $10.2\pm4.2$ \\ 
5.3        &   	4.3 &  $0.29\pm0.21$ &  $0.112\pm0.079$ & $11.2\pm8.5$  \\\hline
5 (average)        	 & 	 &  &  $0.119\pm0.032$ & $11.9\pm3.8$  \\\hline
\hline
\end{tabular}
\end{center}
$^{\rm{a}}$\,From the Limousin et al.~(2016) non-cored lensing model ($z=4.1$) and the Diego et al.~(2015) lensing model ($z=4.3$). For corresponding magnifications, see Table \ref{tab:red}.  \\
$^{\rm{b}}$\,We measure the 1.1$\,$mm flux at the known optical position of each multiple image to mitigate the effects of flux boosting.\\
$^{\rm{c}}$\,Errors include the uncertainty due to the magnification and the photometric uncertainty. \\ 
$^{\rm{d}}$\,Errors include the uncertainty due to the SED template \citep[$27\%$,][]{Kirkpatrick2015}, the magnification and the photometric uncertainty. \\ 
\end{table*}

\subsection{Redshift}
\label{sec:redshift}
The multiple images (5.1, 5.2, 5.3) have independent redshift estimates from blind photometric redshift catalogs (Figure \ref{fig:sed} {\it right}, see also Postman et al.~2012) and from the lensing models (Diego et al.~2015; Limousin et al.~2016, see also Johnson et al.~2014; Zitrin et al.~2015). Table \ref{tab:red} summarizes the redshift estimates for each multiple image. As described in Section \ref{sec:SED}, fitting the optical spectral energy distribution gives photometric redshifts of $\sim4.4$--4.6. The lensing models agree on a redshift of $z\gtrsim4$ for this multiply-imaged source, and are consistent with the $\pm3\sigma$ limits from the photometric redshift. Treu et al.~(2015) suggest a low redshift solution of $z=0.928$ for image 5.1 based on {\it HST} grism data. However, we do not see any strong features in the spectrum and this redshift is not compatible with the mass models of MACSJ0717.5+3745. 

In this paper, we consider two redshift solutions; 1) $z=4.1\pm0.2$ from the non-cored mass model of Limousin et al.~(2016), and 2) $z=4.3$ from the lens model of Diego et al.~(2015). Table \ref{tab:red} lists the magnifications for each of these solutions.  
While a spectroscopic redshift for this multiply-imaged system will be important for further studies, the uncertainty in the analysis in this paper is less affected because of the negative $K$-correction, which makes the relation between millimeter flux and luminosity roughly constant between $z=1$--6.

\begin{table*}
\begin{center}
\scriptsize
\caption{Redshift estimates of the multiply-imaged system 5.1/5.2/5.3 in MACSJ0717.5+3745}
\label{tab:red}
\begin{tabular}{cccccccc}
\hline
ID & RA  & DEC & \multicolumn{2}{c}{Limousin+16 non-cored lensing model} & \multicolumn{2}{c}{Diego+15 lensing model}  & SED fitting \\
 &  &  & z &  magnification$^{\rm{a}}$  & z$^{\rm{b}}$ & magnification$^{\rm{a}}$ & z [$1\sigma$ lower,$1\sigma$ upper] \\\hline
5.1         &   07:17:31.178  & 	+37:44:48.70  & 	$4.1\pm0.2$ &  $6.8\pm1.1$  & 	$4.3$ & $4.5\pm0.6$  &  4.48 [4.21,4.73] \\
5.2      &   07:17:30.698  & 	+37:44:34.12  & 	$4.1\pm0.2$ & $7.5\pm1.0$  & 	$4.3$ & $7.2\pm1.1$  &  4.64 [4.51,4.78] \\
5.3        &   07:17:36.007  & 	+37:46:02.64  & 	$4.1\pm0.2$ &  $3.0\pm0.3$  & 	$4.3$ &  $2.6\pm0.4$ &  4.44 [4.24,4.64]  \\\hline
\end{tabular}
\end{center}
$^{\rm{a}}$\,An additional 10\% uncertainty is added in quadrature to account for differential lensing (see Section \ref{sec:DL}).  \\
$^{\rm{b}}$\,Uncertainties on the redshift are not provided for this lensing model.  \\
\end{table*}

\subsection{UV to near-IR properties}
\label{sec:SED}
Measuring the UV to near-IR photometry for this multiply-imaged system is complicated since the images are extended (Figure \ref{fig:hst} {\it right}) and resolve into separate entries in our multi-wavelength catalog. We take the weighted mean of these entries to estimate the total flux in each band for 5.1, 5.2 and 5.3. Having three lensed images of the same galaxy provides an independent check on the photometry. 

\begin{figure*}[ht!]
\hspace{0.5in}\includegraphics[trim=0 375 0 0,clip,scale=0.7]{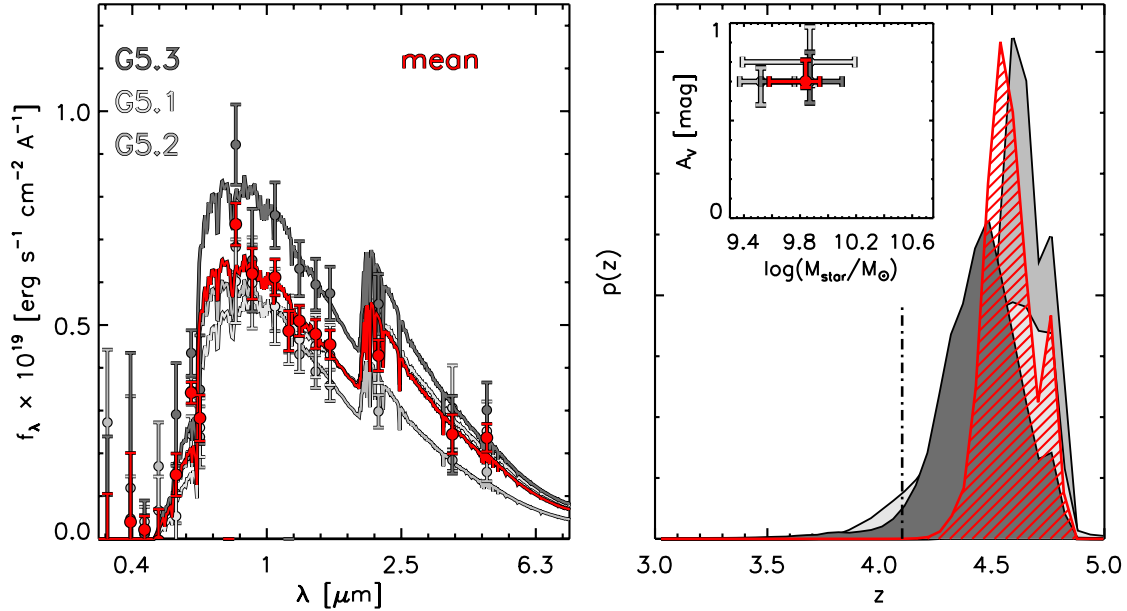}  
\caption{({\it left}) Rest-frame, de-magnified optical spectral energy distribution of 5.1 (light gray), 5.2 (medium gray), 5.3 (dark gray) and the mean of all three sources (red), assuming the redshift and magnifications from the Limousin et al.~(2016) non-cored lensing model. Given the large differences in magnification factors of the three sources, the de-magnified SEDs are remarkably consistent. ({\it right}) Redshift probability distribution from the SED fitting for 5.1 (light gray), 5.2 (medium gray) and 5.3 (dark gray) and for the mean (red). The Limousin et al.~(2016) non-cored solution of z=4.1 (vertical dash-dot line) is consistent with the $3\sigma$ limits of the optical photometric redshift estimates. 
}
\label{fig:sed}
\end{figure*}

We fit the de-magnified UV to the near-IR photometry using FAST (Kriek et al.~2009) adopting Bruzual \& Charlot (2003) stellar populations (BC03), a Chabrier (2003) IMF and a delayed exponentially declining SFH in order to determine the stellar mass. 
Since we find that this galaxy lies closer to the SMC dust curve (see Section \ref{sec:irx}), we perform the SED fitting using SMC dust attenuation and sub-solar metallicity ($Z=0.2\times Z_{\odot}$)\footnote{If we instead assumed a \citet{Calzetti2000} dust attenuation law and a range in metallicities (from sub-solar to super-solar), the stellar masses from the best-fit SEDs are slightly larger, but consistent within the uncertainties, than the values in Table \ref{tab:derived}.}.
The difference in stellar mass between a Chabrier and Kroupa (2001) IMF is negligible (e.g.~Speagle et al.~2014). The uncertainty in the stellar masses include the photometric error, the uncertainty in the magnification (including an additional 10\% for differential lensing, Section \ref{sec:DL}) and the uncertainty in the SED fitting. The stellar masses and their 68\% confidence ranges are given in Table \ref{tab:derived}. Figure \ref{fig:sed} {\it left} shows the SED fits for the $z=4.1$ lens model; given the large magnification values for each multiple-image, the de-magnified SEDs show very good agreement. 

We also fit the de-magnified UV to the near-IR photometry using EAZY (Brammer et al.~2008) to independently determine the photometric redshift. In the {\it right} panel of Figure \ref{fig:sed}, we show the redshift probability distribution for each multiple image and the average of the three. While the redshift solutions from the lensing models are lower, they are consistent within the $\pm3\sigma$ limits of the photometric redshifts from the SED fitting. A spectroscopic redshift for this multiply-imaged source will help further refine the lensing models.

The rest-frame UV spectral slope ($\beta$, where $f_{\lambda}\propto\lambda^{\beta}$) is calculated by fitting a power-law to the rest-frame photometric data between the wavelength range of 1300--3000$\AA$. Prior to fitting, the lensing magnification is removed from the photometric data and we propagate the magnification uncertainty. 
The UV luminosity, $L_{1600}=\nu_{1600}L_{\nu}(1600$\AA), is determined using the fitted value for $\beta$.
Table \ref{tab:derived} lists these derived UV properties corrected for magnification for each multiple image and the average of all three. 

\subsection{Star formation rates}

Our AzTEC detection at 1.1$\,$mm corresponds to a rest wavelength of $<220\,\mu$m at $z>4$, which probes near the peak of the infrared dust emission. At this rest-wavelength, we are most sensitive to the IR luminosity and not the dust mass since we are not in the Rayleigh-Jeans tail of the dust distribution (e.g.~$\lambda_{\rm{rest}}>250\,\mu$m, Scoville et al.~2016). Given the observed faintness of MACS0717\_Az9 at 1.1$\,$mm, we do not expect to detected it with {\it Herschel} (Rawle et al.~2016). 
In order to determine the total IR (8--1000$\,\mu$m) luminosity, $L_{\rm{IR}}$, we must extrapolate using the expected SED for this galaxy. \citet{Kirkpatrick2015} derived representative SED templates from a sample of 343 high redshift galaxies with extensive IR data including mid-IR spectroscopy. 
We fit the intrinsic 1.1$\,$mm flux, after correcting for magnification, to the SED template for a typical high redshift star-forming galaxy \citep{Kirkpatrick2015}. 
The $L_{\rm{IR}}$ of each component of system 5 for the two lensing models are listed in Table \ref{tab:obs}. 

We use the formulas from Murphy et al.~(2011b) to calculate the SFRs (assuming a Kroupa IMF). From $L_{\rm{IR}}$ and $L_{\rm{FUV}}$, we calculate the obscured $\rm{SFR_{IR}}$ and unobscured $\rm{SFR_{UV}}$, respectively. 
The total SFR is then calculated by summing the IR and UV SFRs. 
All star formation rate values are listed in Table \ref{tab:derived}. We find that even though this galaxy has an intrinsically low SFR, at least 75\% of the star formation is obscured. We tested our analysis with different SFR calibrations (e.g.~Calzetti 2013); the obscured fraction is only slightly lower (65\%) and our conclusions are unchanged. 

\begin{table*}
\begin{center}
\scriptsize
\caption{Derived intrinsic physical properties } 
\label{tab:derived}
\begin{tabular}{lllllllllll}
\hline
ID & z$^{\rm{a}}$  &  log(M$_{\ast}$/[M$_{\odot}$]) & $L_{\rm{1600\AA}}$$^{\rm{b}}$ & $\beta$ &  $\rm{SFR_{UV}}$ & $\rm{SFR_{IR}}$  & $\rm{SFR_{total}}$ & $\rm{f_{obscured}}^{\rm{c}}$ \\
&  & [$1\sigma$ lower,$1\sigma$ upper] & ($10^{10}\,L_{\odot}$)  &   & ($\rm{M_{\odot}/yr}$) & ($\rm{M_{\odot}/yr}$) & ($\rm{M_{\odot}/yr}$) & \\\hline
5.1                                & 	4.1 &  9.87 [9.44,10.13]   & $2.03\pm0.24$   & $-0.47\pm0.39$  &     $3.4\pm0.4$  & $14.4\pm6.3$   &     $17.8\pm6.3$   & 0.81  \\
5.2 (MACS0717\_Az9)      & 	4.1 &  9.52 [9.37,9.76]  & $2.29\pm0.23$ & $-0.95\pm0.33$ &      $3.8\pm0.4$   & $14.6\pm5.8$   &     $18.4\pm5.8$   &  0.79    \\
5.3                                 & 	4.1 & 9.87 [9.52,10.08]    &  $2.95\pm0.44$ & $-0.67\pm0.26$ &   $4.9\pm0.7$   & $14.6\pm10.8$   &     $19.5\pm10.8$      & 0.75  \\\hline
5 (average)                             & 	&  9.84 [9.58,9.94]   & $2.43\pm0.18$   & $-0.70\pm0.19$ &   $4.1\pm0.3$  & $14.6\pm4.5$    &     $18.7\pm4.5$   &  0.78\\\hline\hline
5.1                                   & 	4.3 &  10.16 [9.87,10.47]  &   $3.51\pm0.29$ & $-0.48\pm0.26$ &    $5.9\pm0.5$    & $21.6\pm9.3$    &     $27.5\pm9.3$  &   0.79   \\
5.2 (MACS0717\_Az9)      &  4.3 & 9.82 [9.54,9.96]  &  $2.68\pm0.25$  & $-0.95\pm0.30$ &   $4.5\pm0.4$   & $15.3\pm6.3$    &     $19.8\pm6.3$   & 0.77 \\
5.3                                 & 	4.3 & 10.11 [9.72,10.37]  &   $3.76\pm0.56$ & $-0.59\pm0.31$ &   $6.3\pm0.9$  & $16.8\pm12.8$    &     $23.1\pm12.8$    &  0.73  \\\hline
5 (average)                            & 	 & 10.12 [9.82,10.28] & $3.31\pm0.23$   & $-0.67\pm0.17$ &   $5.5\pm0.4$  & $17.9\pm5.7$    &     $23.4\pm5.7$      &  0.76  \\\hline
\hline
\end{tabular}
\end{center}
$^{\rm{a}}$\,From the Limousin et al.~(2016) non-cored lensing model ($z=4.1$) and the Diego et al.~(2015) lensing model ($z=4.3$). For corresponding magnifications, see Table \ref{tab:red}.  \\
$^{\rm{b}}$\, $L_{\rm{FUV}}\sim0.97\times\,L_{\rm{1600\AA}}$ for these galaxies. \\
$^{\rm{b}}$\, $\rm{f_{obscured}}=\rm{SFR_{IR}}$/$\rm{SFR_{total}}$
\end{table*}

\subsection{Differential lensing}
\label{sec:DL}
We are assuming that the magnifications derived from the optical lensing maps also apply to the longer wavelength millimeter data. For highly magnified sources, differential lensing becomes important, where extended and compact regions of a galaxy can be magnified by different factors. Hezaveh et al.~(2012) model the effects of differential lensing for strongly lensed, dusty galaxies. They find that for moderate magnifications similar to MACS0717\_Az9 ($\mu\sim7$), the distribution of flux ratios between the extended and compact regions of a galaxy is peaked at 1 with a FWHM of $\sim0.25$ (i.e.~$\sim10\%$ uncertainty), suggesting that differential lensing is not a large effect. 

In this paper, our main comparison is between the unobscured (UV) and obscured (IR/submm) SFRs. Dusty galaxies have been found to have similar radii of $\sim$2 kpc as measured in UV and (sub)mm images, while the optical sizes which trace the stellar light are more extended (Swinbank et al.~2010; Hodge et al.~2016). 
However, the UV and (sub)mm emission is not always co-spatial and can be offset by up to an arcsec (e.g.~Iono et al.~2006). In order to quantify the range of magnifications that might be applicable to the millimeter emission, we explore a wider area in the non-cored magnification map at $z=4.1$. 
For a lensing magnification of 7.5, 1 arcsecond offset in the source plane corresponds to 2.7Ó in the image plane. Within a 2.7 arcsec diameter circle around the location of the optical lensed galaxy 5.2 (where $\mu=7.5$), we find the magnification ranges from 6.0--9.2 with a standard deviation of 0.76. Therefore, if the UV and millimeter emission is not co-spatial and are magnified by different amounts, this would result in an additional uncertainty of $\sim10\%$ in the magnification, and intrinsic flux, we derive. 

Given these two tests of the effects of differential lensing, we conservatively propagate an additional uncertainty of 10\% in the lensing magnification factors, which is the best we can do until we are able to spatially resolve the dust emission with ALMA.

\section{Discussion}

We have detected dust emission in a strongly-lensed, multiply-imaged galaxy at $z>4$. The high magnification ($\mu=7.5$) predicts that MACS0717\_Az9 has an intrinsic $L_{\rm{IR}}<10^{11}\rm{L_{\odot}}$ ($\rm{SFR_{IR}}=\,14.6\,\rm{M_{\odot}/yr}$). Previous detections of dust in multiple images of lensed galaxies have been limited to ULIRGs at $z<3$ (Borys et al.~2004; Sheth et al.~2014; Kneib et al.~2015). There are very few galaxies at $z>4$ at the low luminosities of MACS0717\_Az9 that have been detected in dust emission (e.g.~Capak et al.~2015; Schaerer et al.~2015; Watson et al.~2015; Willott et al.~2015), and MACS0717\_Az9 provides a unique opportunity to probe the star formation and dust properties in a typical galaxy at this early epoch.

\subsection{IRX-$\beta$}
\label{sec:irx}
\begin{figure*}[ht!]
\includegraphics[trim=70 375 0 75,clip,scale=0.9]{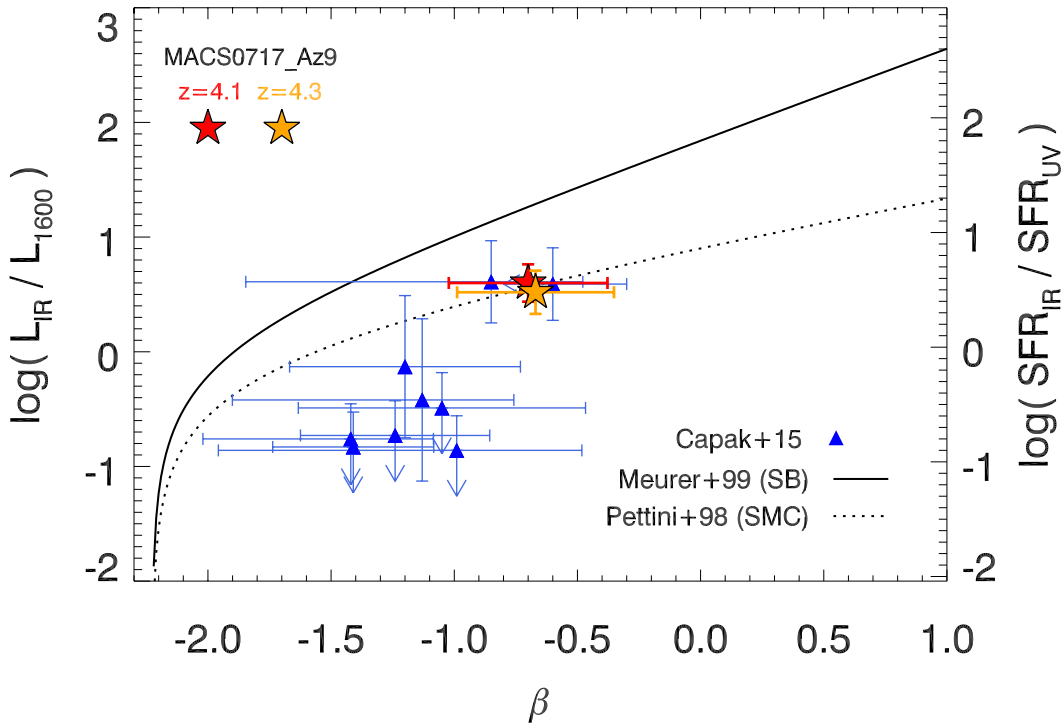}  
\caption{IRX-$\beta$ plot showing the relations for local starburst galaxies (solid curve, Meurer et al.~1999) and the SMC (dotted curve, Pettini et al.~1998). We show MACS0717\_Az9 for the $z=4.1$ and $z=4.3$ lens models as the red and orange stars, respectively.
Even though $75$--80\% of the star formation is coming out in the infrared, we find that MACS0717\_Az9 is consistent with the SMC dust curve, similar to the Capak et al.~(2015) $z\sim5$ UV-selected galaxies (blue triangles).}
\label{fig:irx}
\end{figure*}

UV surveys rely on the UV slope, $\beta$, and its dependence on $\rm{L_{UV}}$ to estimate the dust extinction since IR observations are typically not deep enough. The measured value of $\beta$ for MACS0717\_Az9 is high relative to the distribution of $\beta$ found for UV-selected galaxies of similar luminosity at $z\sim4$ (Bouwens et al.~2012: $\beta_{\rm mean}=-2.01$, $\sigma=0.27$, see also Bouwens et al.~2016); this means that $\rm{L_{UV}}$ alone would underestimate the IR luminosity by an order of magnitude. 

In Figure \ref{fig:irx}, we plot the IRX-$\beta$ relation, which compares the ratio of $\rm{L_{IR}}$/$\rm{L_{UV}}$ to the UV slope $\beta$. 
The solid curve is the established relationship for local starburst galaxies (Meurer et al.~1999) where $z\sim2$ massive UV-selected galaxies are also found (Reddy et al.~2012). The dotted curve shows the milder dust extinction found in the SMC (Pettini et al.~1998). Capak et al.~(2015) found UV-selected galaxies at $z\sim5$ to be closer to this SMC dust curve (see also Murphy et al.~2011a; Lee et al.~2012). Recently, Bouwens et al.~(2016) found that sub-$L^{\ast}$ galaxies also show lower values of IRX, even below the SMC. MACS0717\_Az9 is shown as the red and orange stars, which are closer to the SMC dust curve than the Meurer relation, even though the dust-obscured emission dominates the SFR. 

A lower value in IRX-$\beta$ relative to the starburst relation is usually interpreted to suggest a lower metallicity. It may seem unusual for such a dust-obscured galaxy (75--80\% of star formation is obscured) to have a lower metallicity. Schneider et al.~(2016) recently found significant dust emission in a local, metal-poor dwarf galaxy. By comparing to models of chemical evolution, they conclude that dust content may depend more on the density of the interstellar medium (ISM) than the metallicity, and that in-situ grain growth should be especially important in the early Universe. Future observations of lines sensitive to the ISM density and metallicity such as CO, [CII] and HCN in MACS0717\_Az9 can be used to test this idea.

\begin{figure}[ht!]
\includegraphics[trim=50 0 0 0,clip,scale=0.5]{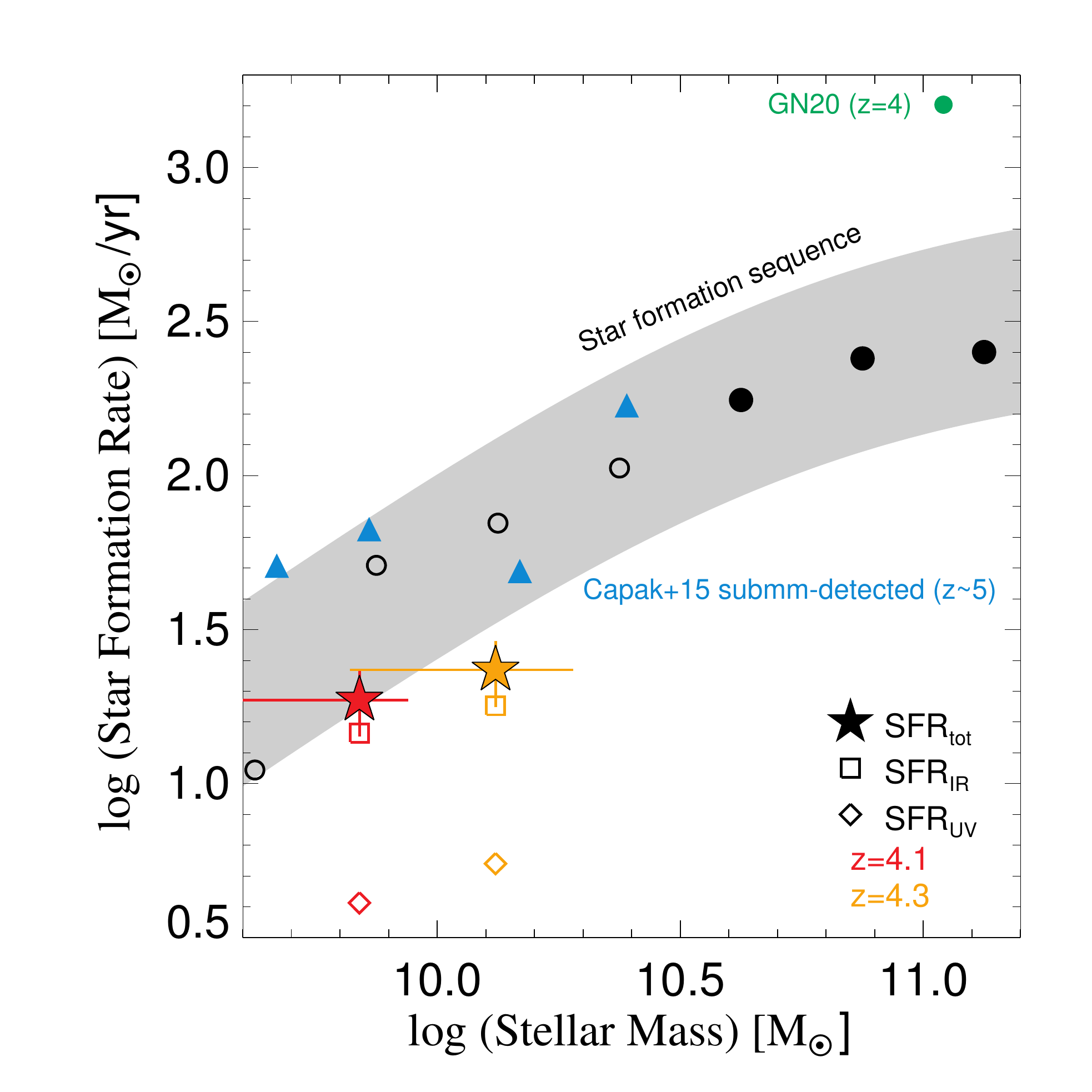}
\caption{The star formation sequence at $z=3$--4 from Tomczak et al.~(2016, solid circles show where the data are mass-complete while open circles are incomplete measurements); the shaded region shows the $\pm1\sigma$ best-fit relation to these data. 
We show the UV (unobscured), IR (obscured) and total SFRs for MACS0717\_Az9 as the red and orange symbols for $z=4.1$ and $z=4.3$, respectively (Table \ref{tab:derived}). Error bars are plotted only for the total SFR points for clarity. 
We overplot the Capak et al.~(2015) UV-selected galaxies that are detected in the submm continuum and the intrinsically luminous submillimeter galaxy GN20 (Pope et al.~2006; Riechers et al.~2014; Tan et al.~2014). All measurements are based on a Chabrier or Kroupa IMF which give similar values for SFR and stellar mass (Speagle et al.~2014). }
\label{fig:ms}
\vspace{0.1in}
\end{figure}

\subsection{Galaxy star formation sequence}

At a given epoch, the tight relationship between the star formation rate and stellar mass implies that most normal star-forming galaxies are undergoing steady growth (e.g.~Noeske et al.~2007; Daddi et al.~2007).  
The stellar mass of MACS0717\_Az9 is well below the estimated knee in the stellar mass function at $z\sim4$ (Muzzin et al.~2013). In order to determine if MACS0717\_Az9 is a normal galaxy for this epoch, we compare the position of this galaxy to the estimated extrapolation of the star formation sequence. In Figure \ref{fig:ms}, we plot the star formation sequence at $z=3$--4 from Tomczak et al.~(2016). 
The $z\sim5$ star-forming galaxies detected in submm continuum with ALMA by Capak et al. (2015) and the extreme submillimeter galaxy, GN20 (Pope et al.~2006; Riechers et al.~2014; Tan et al.~2014), are shown for comparison.

First, we find that MACS0717\_Az9 (red and orange stars for two redshift solutions) is consistent with the estimated star formation sequence for this epoch, and that this galaxy resides in a region that is relatively unexplored in the infrared. Given the error bars on MACS0717\_Az9 and the uncertainty in the star formation sequence at these low masses, we do not claim MACS0717\_Az9 is below the star formation sequence but we can confidently say that the source is not an extreme starburst galaxy like GN20. 

Second, we show that even though this is a normal star-forming galaxy, its SFR is dominated by the obscured component ($\rm{SFR_{IR}}$ is at least 75\% of the total SFR). This underscores the importance of accurately including this obscured component when accounting for the global SFRD, even at these high redshifts and lower stellar masses, and stresses the need for deep and wide IR/submm surveys. 

Two recent papers that surveyed the Hubble Ultra Deep Field with ALMA seem to suggest a smaller number of high redshift galaxies detected in dust emission than expected (Dunlop et al.~2017; Bouwens et al.~2016). Without lensing, a galaxy like MACS0717\_Az9 would not have been detected at the depth of the Dunlop et al.~(2017) ALMA map. The one $z>3.5$ galaxy detected in the Dunlop et al.~(2017) observations has a slightly lower stellar mass ($4\times10^{9}\rm{M_{\odot}}$) and higher $\rm{SFR_{IR}}=37\,\rm{M_{\odot}/yr}$ than MACS0717\_Az9, but is similarly dominated by the dust-obscured star formation ($\rm{f_{obscured}}=0.94$). 
The high levels of dust obscuration observed in a handful of normal galaxies at $z>4$ suggests that we cannot easily rule out the importance of dust emission in galaxies at $z>4$.

\subsection{What if 5.2 only has half the millimeter flux?}
\label{sec:whatif}

In the previous sections, we assumed that all of the millimeter flux from MACS0717\_Az9 is associated with 5.2 since that is the most likely result from our statistical analysis (Figure \ref{fig:Pval}). Here we explore how our results are affected if only half the flux of MACS0717\_Az9 is associated with 5.2. If the millimeter flux is half what we assumed in Table \ref{tab:obs}, then $\rm{L_{IR}}=4.9\times10^{10}\,L_{\odot}$, $\rm{SFR_{IR}}=7.3\,\rm{M_{\odot}/yr}$, $\rm{SFR_{total}}=11.4\,\rm{M_{\odot}/yr}$ and $\rm{f_{obscured}}=0.64$, assuming the $z=4.1$ lens model. Under this assumption, the obscured SFR is slightly lower, but the total SFR is still dominated by the dust-obscured contribution. 
This places MACS0717\_Az9 even lower on both the star formation sequence (Figure \ref{fig:ms}) and the IRX-$\beta$ plot (Figure \ref{fig:irx}). Therefore, the results and implications discussed in this paper are unchanged. Future observations with ALMA will confirm how much millimeter flux is associated with this multiply-imaged source.

\section{Summary}
\label{sec:sum}

We have directly detected dust emission in an intrinsically lower-luminosity ($\rm{L_{IR}}=9.7\times10^{10}\,L_{\odot}$) galaxy at $z>4$ with AzTEC on the LMT. Currently, this is the only star forming galaxy at such a low luminosity (sub-LIRG) where multiple images are detected in dust emission.
While the SNRs of the individual images are modest, the false detection rate for randomly detecting two multiple images of a known system at the correct flux ratio given the known magnifications is negligible. We calculate the unobscured SFR from the UV and the obscured SFR from the IR and calculate a total intrinsic SFR of $18.7\,\rm{M_{\odot}/yr}$, 75--80\% of which is obscured. MACS0717\_Az9 is a normal star-forming galaxy with an intrinsic stellar mass of $6.9\times10^{9}\rm{M_{\odot}}$ and is consistent with the estimated star formation sequence at $z\sim4$. The dust obscuration in MACS0717\_Az9 appears to be more like that of the SMC than local starburst galaxies. While we might expect lower metallicities for a lower mass galaxy 1.5$\,$Gyrs after the Big Bang \citep{RemyRuyer2015}, our rest-frame IR observations find a significant dust component. Further observations to constrain the conditions of the the gas and dust in MACS0717\_Az9 and future surveys with the 50m LMT (observational limit of $\rm{L_{IR}}\sim10^{11}\,L_{\odot}$) will help constrain the buildup of metals and dust in early galaxy evolution.

\acknowledgments
We thank the referee for constructive comments on this paper. 
This  work  would  not  have  been  possible  without the  long-term  financial  support  from the Mexican  Science  and Technology  Funding  Agency,  CONACYT during  the  construction  and  early  operational phase of the Large Millimeter Telescope Alfonso Serrano, as  well  as  support  from  the  the US National  Science  Foundation (NSF) via the University Radio Observatory program, INAOE  and the University of Massachusetts, Amherst.
This work utilizes gravitational lensing models produced by PIs Bradac, Natarajan \& Kneib (CATS), Merten \& Zitrin, Sharon, and Williams, and the GLAFIC and Diego groups. This lens modeling was partially funded by the HST Frontier Fields program conducted by STScI. STScI is operated by the Association of Universities for Research in Astronomy, Inc. under NASA contract NAS 5-26555. The lens models were obtained from the Mikulski Archive for Space Telescopes (MAST). We also thank J.~Diego for information on their lens model. 
This work is based in part on observations made with the {\it Spitzer Space Telescope}, which is operated by the Jet Propulsion Laboratory, California Institute of Technology under a contract with NASA.
AP  acknowledges  valuable  discussions  with Sandy Faber, George Rieke and Adi Zitrin. 
ML acknowledges CNRS and CNES for support.
IA and DRG acknowledge support from CONACYT grants CB-2011-01-1672 and CB-2011-01-167281, respectively. 
ML acknowledges the Centre National de la Recherche Scientifique for support.
DM acknowledges the support of the NSF under Grant 1513473, and from HST-AR-14302, provided by NASA through a grant from STScI, which is operated by AURA, under NASA contract NAS5-26555. DLV acknowledges support from the NASA Keck PI Data Awards, administered by the NASA Exoplanet Science Institute.
KEW gratefully acknowledge support by NASA through Hubble Fellowship grant \#HF2-51368 awarded by the Space Telescope Science Institute, which is operated by the Association of Universities for Research in Astronomy, Inc., for NASA.

\vspace{5mm}
\facilities{LMT, HST, Spitzer}

\listofchanges

\end{document}